\documentclass[draftclsnofoot,onecolumn]{IEEEtran}
\usepackage{amsmath,graphicx}
\usepackage{amsthm,amssymb,cite}
\usepackage{eucal}
\usepackage{xifthen}
\usepackage{mathtools}
\usepackage{enumerate}
\usepackage{microtype}
\usepackage{xspace}
\usepackage{bm}

\theoremstyle{plain}
\newtheorem{theorem}{Theorem}

\theoremstyle{definition}

\newtheorem{example}{Example}

\title{Random Sampling for Distributed Coded Matrix Multiplication}
\author{
Wei-Ting~Chang \qquad Ravi~Tandon\\ 
Department of Electrical and Computer Engineering\\
University of Arizona, Tucson, AZ, USA\\
E-mail: \{\textit{wchang, tandonr}\}@email.arizona.edu
}

\begin{document}
%
\maketitle
\begin{abstract}
Matrix multiplication is a fundamental building block for  large scale computations arising in various applications, including machine learning. There has been significant recent interest in using coding to speed up distributed matrix multiplication, that are robust to stragglers (i.e., machines that may perform slower computations). In many scenarios, instead of exact computation, approximate matrix multiplication, i.e., allowing for a tolerable error is also sufficient. Such approximate schemes make use of randomization techniques to speed up the computation process. In this paper, we initiate the study of approximate coded matrix multiplication, and investigate the joint synergies offered by randomization and coding. Specifically, we propose two coded randomized sampling schemes that use (a) codes to achieve a desired recovery threshold and (b) random sampling to obtain approximation of the matrix multiplication. Tradeoffs between the recovery threshold and approximation error obtained through random sampling are investigated for a class of coded matrix multiplication schemes. 

\noindent \textit{\textbf{Keywords --} Matrix multiplication, Random sampling, Coded Distributed Computing}
\end{abstract}
%
%
\section{Introduction}
\label{sec:intro}
\footnote{This work was supported by NSF Grant CAREER 1651492.}
Matrix multiplication has been one of the most essential fundamental building blocks for various applications in fields such as signal and image processing, machine learning, optimization and wireless communications. Outsourcing the computations to distributed machines has become a preferable way to speed up the process when one is dealing with large scale data. However, distributed systems suffer from the straggler effect where the slowest worker(s) can limit the speed-ups offered by distributed computation.

In order to mitigate the impact of stragglers, the idea of using coded distributed computation has gained significant recent interest. In general, these codes are used to introduce redundancy to the computations. For example, by applying one of the simplest codes - repetition codes, one can let multiple machines work on the same computation. One can then obtain the desired result whenever the fastest machine finishes the assigned tasks. Much more efficient codes have been applied to the distributed computing problems. Significant recent progress has been made on understanding the additional speed-ups gained by mitigating stragglers using codes. Several codes that are particularly efficient for the distributed matrix multiplication problems include Polynomial codes, MatDot codes and Lagrange codes \cite{PolyCode2017,PolyCode2018,MatDot,Lagrange2018}. These codes add redundancy in a way that one can obtain the desired result with the responses from an arbitrary subset of machines. The smallest number of machines which allow perfect recovery of the computation is referred as the recovery threshold. 

In contrast to adding redundancy, another methodology to speed up matrix multiplication comes from the idea of randomization. By allowing some tolerable error in the computation, randomized algorithms can provide speed-ups by working on matrices of smaller dimensionality. However, the randomization techniques must be carefully designed, in order to provide guarantees on the error. 
Random sampling and random projection are two commonly used techniques for this purpose. Random sampling algorithms sample either the columns or rows from the original matrix to construct sketches of original matrices, and the subsequent task is performed on sketched matrices. The key to a good sampling scheme is to carefully design what to sample, since not all columns/rows carry the same amount of information. Several works on random sampling include \cite{DriKanMah2006,DesRadLui2006,BouMahDri2008,GurSin2011,BouDriMag2014,BouWoo2017}. Random projection algorithms construct the sketch matrix by projecting the original matrix to a vector space with a lower dimension. Projection algorithms are typically designed to have good distance preserving properties (Johnson-Lindenstrauss lemma \cite{Ach2003, DG2003}), and have been investigated in various works \cite{Ach2003,DG2003,Sarlos2006,AilCha2009,ClaWoo2013,CohNelWoo2015}.

\textbf{\textit{Main Contributions: }} In this paper, we explore the synergies between coding and randomization, and explore the tradeoffs between reconstruction error and  recovery threshold for distributed matrix multiplication. To answer this question, we devise two novel coded sampling schemes that can achieve various levels of speed-ups depending on how well one wishes to approximate the desired result. For the scope of this paper, we focus on Matdot codes \cite{MatDot}, and design sampling strategies tailored to these codes. We present a family of coded sampling schemes, which sample a sub-set of columns from the matrices, followed by application of Matdot codes on the sampled matrices. We analyze two sampling strategies: one where the sampling of rows/columns is done independently (with replacement), and one where we sample a subset of rows/columns (without replacement). 

We show that if the matrices $A, B$ to be multiplied are divided into $m$ parts (for details, see Section \ref{sec:RCMM}), and for any integer $1\leq s\leq m$, a recovery threshold of $K=2s-1$ is achievable. Moreover, the expected approximation errors of the proposed coded sampling schemes for a recovery threshold of $K=2s-1$ are as follows: $(a)~\mathbb{E}[\Vert AB - \hat{A}\hat{B}_\mathcal{S}\Vert_F^2 ] = (\sum_{\mathcal{S}}\Vert\sum_{q\in\mathcal{S}} A_q B_q\Vert_F)^2/c^2 - \Vert AB \Vert_F^2$, where $\mathcal{S},~\vert\mathcal{S}\vert=s$ denotes the set of $s$ sampled indices and $c=\binom{m}{s}\cdot s/m$ when coded set-wise sampling scheme is used; and $(b)~\mathbb{E}[\Vert AB - \hat{A}\hat{B}\Vert_F^2] = (\sum_{q=0}^{m-1}\Vert A_qB_q\Vert_F)^2/s - \Vert AB\Vert_F^2/s$ when coded independent sampling scheme is used. These results reveal a tradeoff between recovery threshold and approximation error, i.e., a lower recovery threshold can be obtained by allowing reconstruction error. 

\section{System Model \label{sec:sysModel}}

We consider a distributed system which consists of a master and $N$ workers. Each worker is connected to the master through a separate link. The goal of the master is to approximate matrix multiplication $AB$, where $A\in\mathbb{F}^{d_1\times d_2}$ and $B\in\mathbb{F}^{d_2\times d_3}$, using $N$ workers, in the presence of stragglers, for some sufficiently large field $\mathbb{F}$. We note that depending on the computation strategy used, the master may not need to wait for all $N$ workers to recover the approximation of $AB$. The smallest number of workers needed to recover the approximation is referred as the recovery threshold $K$.

To tolerate stragglers, the master encodes $A$ and $B$ separately, and workers multiply the encoded versions of $A$ and $B$. The encoding functions used are $\bm{f}=(\mathit{f}_0,\cdots,\mathit{f}_{N-1})$ and $\bm{g}=(\mathit{g}_0,\cdots,\mathit{g}_{N-1})$, where $\mathit{f}_n$ and $\mathit{g}_n$ are the encoding functions for worker $n$. Specifically, the encoded matrices for worker $n$ are $\widetilde{A}_n$ and $\widetilde{B}_n$, where $\widetilde{A}_n=\mathit{f}_n(A)$ and $\widetilde{B}_n=\mathit{g}_n(B)$. We denote the answer from worker $n$ as $Z_n=\widetilde{A}_n\widetilde{B}_n$. The master must be able to decode the desired result from any $K$ workers. We denote the approximated result as $\hat{A}\hat{B}=d(Z_{n_0},\cdots,Z_{n_{K-1}})$, where $d(\cdot)$ is the decoding function. The performance of coded sampling schemes is measured through the expected approximation error $\mathbb{E}[\Vert AB - \hat{A}\hat{B} \Vert_F^2]$, where $\Vert M\Vert_F$ denotes the Frobenius norm of a matrix $M$. Note that we choose Frobenius norm for its properties, which will be useful for our analysis. Other norms could potentially be used for evaluating the schemes.


\section{Coded Matrix Multiplication}
\label{sec:matdot}

For the scope of this paper, we focus on one of the codes, namely MatDot codes \cite{MatDot}\footnotemark. We show the intuition behind MatDot codes and its application to approximate matrix multiplication through an illustrative example. 
\begin{example}
Consider a matrix multiplication problem with $N$ workers using $m=2$-MatDot code, where $N\geq 3$. The input matrices are partitioned into $m=2$ submatrices as follows,
\begin{align}
A = \begin{bmatrix}
A_0 & A_1
\end{bmatrix},
\quad
B = \begin{bmatrix}
B_0\\
B_1
\end{bmatrix}, \label{eq:partition}
\end{align}
where $A_q\in\mathbb{F}^{d_1\times \frac{d_2}{2}}$ and $B_q\in\mathbb{F}^{\frac{d_2}{2}\times d_3},\text{ for }q=0,1$. The product of $AB$ can then be written as,
\begin{align}
AB = A_0B_0 + A_1B_1. \label{eq:sum}
\end{align}
The submatrices $A_q$ and $B_q$ are encoded as follows,
\begin{align}
\widetilde{A}_n = A_0 + x_nA_1,\quad\widetilde{B}_n = x_nB_0 + B_1,
\end{align}
for $n=0,\cdots,N-1$, where $\widetilde{A}_n$ and $\widetilde{B}_n$ have the same dimensions as $A_q$ and $B_q$, and $x_n \in \mathbb{F}$ is a distinct non-zero element  assigned to worker $n$. After encoding, worker $n$ computes $\widetilde{A}_n\widetilde{B}_n$ and sends the result to the master. Without loss of generality, we assume that the first $3$ workers respond and the master receives,
\begin{align}
Z_0 &= \widetilde{A}_0\widetilde{B}_0 = A_0B_1 + (A_0B_0+A_1B_1)x_0 + A_1B_0x_0^2,\nonumber\\
Z_1 &= \widetilde{A}_1\widetilde{B}_1 = A_0B_1 + (A_0B_0+A_1B_1)x_1 + A_1B_0x_1^2,\nonumber\\
Z_2 &= \widetilde{A}_2\widetilde{B}_2 = A_0B_1 + (A_0B_0+A_1B_1)x_2 + A_1B_0x_2^2.\nonumber
\end{align}
It can be seen that the results can be viewed as $3$ distinct evaluations of a degree $2$ polynomial. Thus, the master can apply any polynomial interpolation technique and obtain the coefficients $A_0B_1,A_0B_0+A_1B_1$ and $A_1B_0$ using any $3$ evaluations received. Since the desired result $A_0B_0+A_1B_1$ can be obtained from any $K=3$ evaluations, we say $2$-MatDot code achieves a recovery threshold of $K=3$. 
\end{example}
\footnotetext{We note that there are many other codes that could potentially be applied to our problem, such as Polynomial and Lagrange codes \cite{PolyCode2017,PolyCode2018,Lagrange2018}. Investigating randomization schemes for other codes is part of our ongoing work.}
We now introduce the idea of randomization in this context. In particular, for scenarios where approximate matrix multiplication is sufficient, we show that the recovery threshold can be even reduced to $1$. Using the same partition as the previous example, 
if we want the recovery threshold to be $K=1$, the master can follow the following strategy: it samples one of the submatrices of $A$ and $B$ (i.e., either $(A_0, B_0)$ or $(A_1, B_1)$ with a certain probability). The chosen index is a Bernoulli random variable $Y$. It then assigns each worker to compute $A_Y B_Y$. It waits for only $K=1$ worker, and declares $A_Y B_Y$ as the approximate answer for $AB$. It can be readily shown that the expected value of $A_YB_Y$ is $AB$ with proper scaling. Although $A_YB_Y$ is an unbiased estimator of $AB$ on average, there will be some error in practice, and the sampling scheme must be designed to (a) give an unbiased estimate of $AB$, and (b) minimize the resulting error as much as possible. We first briefly summarize the general construction of MatDot, followed by the details of our randomized sampling scheme.

To apply MatDot codes for any $m$ that divides $d_2$, the input matrices $A$ and $B$ are partitioned into $m$ disjoint submatrices horizontally and vertically, respectively, i.e.,
$A =[A_0 \cdots A_{m-1}],~B = [B_0^T \cdots B_{m-1}^T]^T,$ where $A_q\in\mathbb{F}^{d_1\times \frac{d_2}{m}}$ and $B_q\in\mathbb{F}^{\frac{d_2}{m}\times d_3},~ q=0,\cdots,m-1$. The submatrices of $A$ and $B$ are encoded into $\widetilde{A}_n = \sum_{q=0}^{m-1} A_q x_n^{q},~ \widetilde{B}_n = \sum_{r=0}^{m-1} B_r x_n^{m-1-r}$ for worker $n$, where $x_n$ is a distinct non-zero element in $\mathbb{F}$ assigned to worker $n$. Workers compute the product of their respective $\widetilde{A}_n$ and $\widetilde{B}_n$, and return the results to the master. The results can be seen as a polynomial evaluated at $N$ distinct points, i.e., $h(x)=\sum_{q=0}^{m-1}\sum_{r=0}^{m-1} A_qB_r x^{q+m-1-r}$, where $x=x_n,~n=0,\cdots,N-1$. The degree of this polynomial is $2m-2$, hence, the coefficients of the polynomial can be interpolated using any $2m-1$ evaluations. Note that the desired result is the sum of $A_qB_r,~q=r$, and it is the coefficient of $x^{m-1}$. With the ability of computing the desired result from any $2m-1$ workers, we say $m$-MatDot achieves a recovery threshold of $K=2m-1$ (see \cite{MatDot} for details).

\section{Coded Sampling for Approximate Matrix Multiplication} \label{sec:RCMM}
In this section, we present two coded sampling schemes and study the tradeoff between recovery threshold and approximation error. To apply MatDot, matrices $A$ and $B$ are partitioned into $m$ submatrices horizontally and vertically, respectively. Both schemes sample $s$ submatrices from $A$ and the corresponding submatrices from $B$, and encode them using MatDot, where the choice of $s$ controls both the approximation error and the recovery threshold.

\subsection{Coded Set-wise Sampling \label{sec:setwise}}


For the coded set-wise sampling scheme, the master samples a subset $\mathcal{S}\subset\{0,\cdots,m-1\}$ of the indices of submatrices, where $\vert\mathcal{S}\vert=s\leq m$ is picked according to probability $P_\mathcal{S}$. We denote the sampled submatrices as $A_{\mathcal{S}}\triangleq(A_{q_0},\cdots,A_{q_{s-1}})$ and $B_{\mathcal{S}}\triangleq(B_{q_0},\cdots,B_{q_{s-1}})$. The sampled submatrices are then encoded as,
\begin{align}
\widetilde{A}_n=\sum\limits_{\ell=0}^{s-1}\frac{ A_{q_\ell}x_n^{\ell}}{\sqrt{cP_\mathcal{S}}},\quad \widetilde{B}_n=\sum\limits_{\ell^\prime=0}^{s-1}\frac{B_{q_{\ell^\prime}}x_n^{s-1-\ell^\prime}}{\sqrt{cP_\mathcal{S}}},
\end{align}
where the scaling is done to ensure that the approximation is an unbiased estimator of $AB$ and the choice of the constant $c=\binom{m}{s}\cdot s/m$ will become clear in the analysis. The goal is to approximate $AB$ using the sum of $A_{q_\ell}B_{q_{\ell^\prime}},~\ell={\ell^\prime}=0,\cdots,s-1$. Note that this sum is originally a part of $AB$. Workers are assigned to compute their respective $\widetilde{A}_n\widetilde{B}_n$ and return the results. The master receives the results,
\begin{align}
h(x_{n_k})=\frac{1}{cP_\mathcal{S}}\sum\limits_{\ell=0}^{s-1}\sum\limits_{\ell^\prime=0}^{s-1}A_{q_\ell}B_{q_{\ell^\prime}}x_{n_k}^{\ell+s-1-\ell^\prime},
\end{align}
for $k=0,\ldots, K-1$, corresponding to any $K$ workers. As shown in Section \ref{sec:matdot}, since the degree of this polynomial is $2s-2$, the coefficients of the polynomial can be interpolated using the results from any $K=2s-1$ workers. The master can then obtain the approximation $\hat{A}\hat{B}_\mathcal{S}=\sum_{\ell=0}^{s-1}\sum_{\ell^\prime=\ell}A_{q_\ell}B_{q_{\ell^\prime}}/cP_{\mathcal{S}}$.

Our main result is stated in the following Theorem:
\begin{theorem}
For an approximate coded matrix multiplication problem, to achieve a recovery threshold of $K=2s-1$ using $s$-MatDot codes, the expected approximation error of the coded set-wise sampling scheme is as follows,
\begin{align}
\mathbb{E}\left[\Vert AB - \hat{A}\hat{B}_\mathcal{S}\Vert_F^2 \right] = \frac{\left(\sum\limits_{\mathcal{S}}\Vert\sum\limits_{q\in\mathcal{S}} A_q B_q\Vert_F\right)^2}{c^2} - \Vert AB \Vert_F^2, \nonumber
\end{align}
by sampling using the optimal distribution $P_\mathcal{S}^\star$ shown in the analysis, where $\mathcal{S},~\vert\mathcal{S}\vert=s$ denotes the set of sampled indices and $c=\binom{m}{s}\cdot s/m$.
\label{thm:setwise}
\end{theorem}

To prove Theorem \ref{thm:setwise}, we first show that the approximation $\hat{A}\hat{B}_\mathcal{S}$ is an unbiased estimator of $AB$. We start by looking at the expected value of the $ij$th element of the approximation:
\begin{align}
\mathbb{E}\left[(\hat{A}\hat{B}_\mathcal{S})_{ij}\right]&=\mathbb{E}\left[\sum\limits_{q\in\mathcal{S}}\frac{(A_qB_q)_{ij}}{cP_\mathcal{S}}\right]\nonumber\\
&= \frac{1}{c}\sum\limits_{\mathcal{S}}P_\mathcal{S}\sum\limits_{q\in\mathcal{S}}\frac{(A_qB_q)_{ij}}{P_\mathcal{S}}\\
&=(AB)_{ij}, \label{eq:unbiased}
\end{align}
where \eqref{eq:unbiased} follows from the definition of expected value and the design of the scheme, and $c$ is the number of times each $A_qB_q$ appears in the summation. Thus,
\begin{align}
\mathbb{E}\left[(\hat{A}\hat{B}_\mathcal{S})_{ij}^2\right] = \frac{1}{c^2}\sum\limits_{\mathcal{S}}\frac{(\sum\limits_{q\in\mathcal{S}} A_qB_q)_{ij}^2}{P_\mathcal{S}}.
\end{align}
Since $\text{Var}[(\hat{A}\hat{B}_\mathcal{S})_{ij}]=\mathbb{E}[(\hat{A}\hat{B}_\mathcal{S})_{ij}^2] - \mathbb{E}[(\hat{A}\hat{B}_\mathcal{S})_{ij}]^2$, we have
\begin{align}
\text{Var}\left[(\hat{A}\hat{B}_\mathcal{S})_{ij}\right] = \frac{1}{c^2}\sum\limits_{\mathcal{S}}\frac{(\sum\limits_{q\in\mathcal{S}} A_qB_q)_{ij}^2}{P_\mathcal{S}} - (AB)_{ij}^2.
\end{align}
We next find the expected approximation error by calculating:
\begin{align}
\mathbb{E}\left[\Vert AB - \hat{A}\hat{B}_\mathcal{S}\Vert_F^2 \right] =& \sum\limits_{i=0}^{d_1-1}\sum\limits_{j=0}^{d_3-1}\mathbb{E}\left[(AB - \hat{A}\hat{B}_\mathcal{S})_{ij}^2 \right]\\ 
=& \sum\limits_{i=0}^{d_1-1}\sum\limits_{j=0}^{d_3-1} \text{Var}\left[(\hat{A}\hat{B}_\mathcal{S})_{ij}\right]\nonumber\\
=& \sum\limits_{i=0}^{d_1-1}\sum\limits_{j=0}^{d_3-1} \frac{1}{c^2}\sum\limits_{\mathcal{S}}\frac{(\sum\limits_{q\in\mathcal{S}}A_qB_q)_{ij}^2}{P_\mathcal{S}} - \sum\limits_{i=0}^{d_1-1}\sum\limits_{j=0}^{d_3-1} (AB)_{ij}^2\\
=& \frac{1}{c^2}\sum\limits_{\mathcal{S}}\frac{\Vert\sum\limits_{q\in\mathcal{S}} A_qB_q\Vert_F^2}{P_\mathcal{S}} - \Vert AB\Vert_F^2,\label{eq:errorSet}
\end{align}
where \eqref{eq:errorSet} follows from placing the double summations before $(\sum_{q\in\mathcal{S}} A_qB_q)_{ij}^2$. 

Note that $\Vert AB\Vert_F^2$ is a constant for fixed $A$ and $B$, hence, we can use the method of Lagrange multipliers to find the optimal $P_\mathcal{S}$ by putting $\sum_\mathcal{S} P_\mathcal{S}=1$ as a constraint on the first term in \eqref{eq:errorSet} and solve for the $P_\mathcal{S}$ that minimizes the error. The optimal $P_\mathcal{S}^\star$ can be found to be $P_\mathcal{S}^\star=\Vert\sum_{q\in\mathcal{S}} A_q B_q\Vert_F/\sum_{\mathcal{S'}}\Vert\sum_{q\in\mathcal{S'}} A_q B_q\Vert_F$. Plugging $P_\mathcal{S}^\star$ in \eqref{eq:errorSet} completes the proof of Theorem \ref{thm:setwise}.

We note that the computational complexity of finding the optimal probabilities is $\binom{m}{s}\times O(d_1d_2d_3s/m)$, which can be high. A way to overcome this issue is to sample $A$ and $B$ using uniform distribution $P_\mathcal{S}=1/\binom{m}{s}$ at the cost of higher approximation error. We next propose another alternative (and simpler) sampling strategy and obtain the corresponding approximation error. 

\subsection{Coded Independent Sampling \label{sec:independent}}

For coded independent sampling, at each iteration, the master samples an index $q_t\in [0:m-1]$ according to probability $P_{q_t}$, the probability that $A_{q_t}$ and $B_{q_t}$ being sampled at time $t,~t=0,\cdots,s-1$. After sampling $s$ indices, the corresponding submatrices are encoded into $\widetilde{A}_n=\sum_{t=0}^{s-1} A_{q_t}x_n^{t}/\sqrt{sP_{q_t}},~ \widetilde{B}_n=\sum_{t'=0}^{s-1}B_{q_{t'}}x_n^{s-1-t'}/\sqrt{sP_{q_{t'}}}$. Workers are assigned to compute their respective $\widetilde{A}_n\widetilde{B}_n$. The results the master received are 
\begin{align}
h(x)=\sum\limits_{t=0}^{s-1}\sum\limits_{t'=0}^{s-1} \frac{A_{q_t}B_{q_{t'}}  x_n^{t+s-1-t'} }{s\sqrt{P_{q_t}P_{q_{t'}}}}\nonumber, 
\end{align}
where $x=x_n,~n=0,\cdots,N-1$. The degree of this polynomial is $2s-2$, hence, the coefficients of the polynomial can be interpolated by using the results from any $2s-1$ workers. The master can thus obtain the approximation $\hat{A}\hat{B}=\sum_{t=0}^{s-1}\sum_{t'=t}A_{q_t}B_{q_{t'}}/s\sqrt{P_{q_t}P_{q_t'}}$. The expected error is (following similar steps as in previous section) as follows:
\begin{align}
\mathbb{E}\left[\Vert AB - \hat{A}\hat{B}\Vert_F^2 \right] = \frac{1}{s}\left(\sum\limits_{q=0}^{m-1}\Vert A_qB_q\Vert_F\right)^2 - \frac{1}{s}\Vert AB\Vert_F^2. \nonumber
\end{align}
\begin{figure}[t]
    \centering
    \begin{minipage}[b]{0.48\textwidth}
    \includegraphics[width=0.95\textwidth]{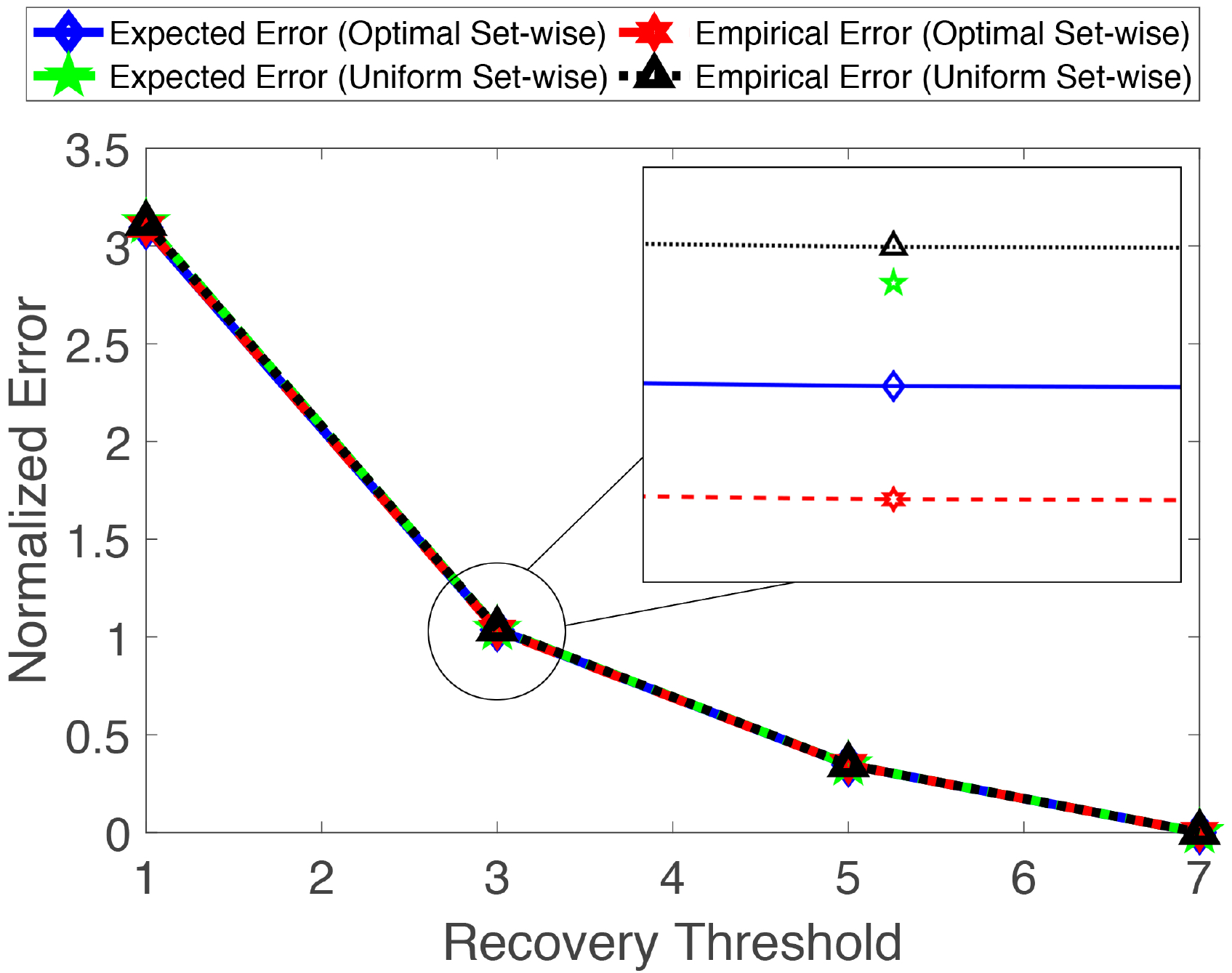}
    \caption{Normalized error for coded set-wise sampling scheme as function of recovery threshold $K$ (errors for $K=3$ are zoomed in).}
    \label{fig:setwise}
\end{minipage}
  \hfill
  \begin{minipage}[b]{0.48\textwidth}
    \includegraphics[width=0.95\textwidth]{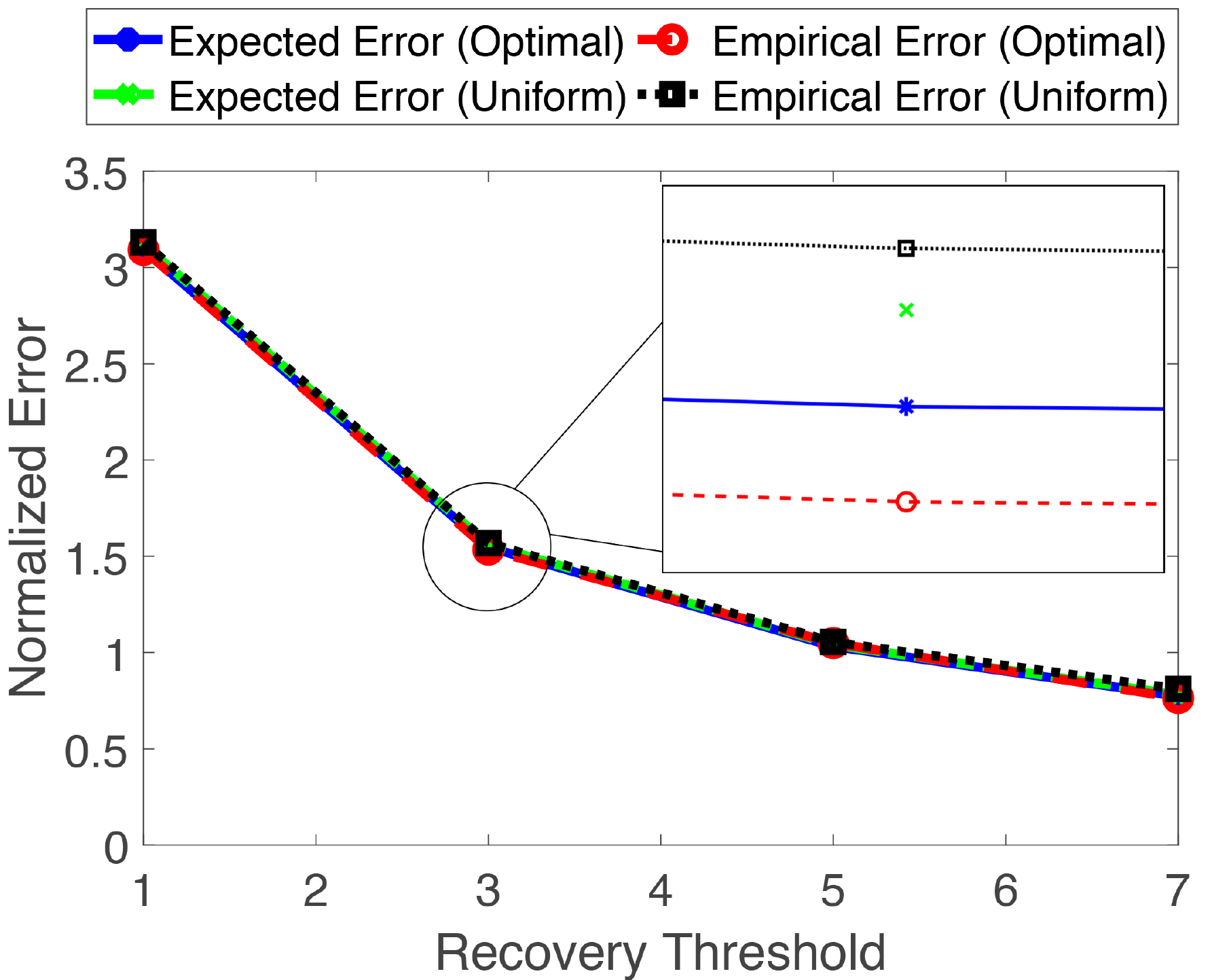}
    \caption{Normalized error for coded independent sampling scheme as function of recovery threshold $K$ (errors for $K=3$ are zoomed in).}
    \label{fig:independent}
    \end{minipage}
\end{figure}

\begin{table}[t]
  \begin{center}
    \begin{tabular}{|c|c|c|c|c|}
    \hline
    	 & \multicolumn{2}{|c|}{\small Independent Sampling} & \multicolumn{2}{|c|}{\small Set-wise Sampling}\\
    	\hline
      \small {Recovery} & \small{Uniform} & \small{Optimal} & \small{Uniform} & \small{Optimal}\\
      \small{Threshold} & & & & \\
      \hline
      $K=1$ & $3.1314$ & $\mathbf{3.0917}$ & $3.1155$ & $3.0972$\\
      $K=3$ & $1.5679$ & $1.5349$ & $1.0409$ & $\mathbf{1.0337}$\\
      $K=5$ & $1.0545$ & $1.0489$ & $0.3468$ & $\mathbf{0.3463}$\\
      $K=7$ & $0.8105$ & $0.7633$ & $\mathbf{0}$ & $\mathbf{0}$\\
	\hline
    \end{tabular}
  \end{center}
  \caption{The normalized empirical errors, where the bolded values indicates the best scheme for each $K$. \label{table1}}
\end{table}

\subsection{Simulation Results}
\label{sec:simResults}
In this section, we present simulation results to show the performance of the two coded randomized sampling schemes. 
We consider the case where $A\in\mathbb{F}^{60\times 4}$ and $B\in\mathbb{F}^{4\times 60}$, where $A$ and $B$ are partitioned into $m=4$ submatrices. With $m=4$, the master can sample either $s=1,2,3$ or $s=4$ submatrices and achieved recovery thresholds of $K=1,3,5$ or $K=7$, respectively. The normalized errors shown in Fig. \ref{fig:setwise}, \ref{fig:independent} and Table \ref{table1} are calculated by computing $\Vert AB-\hat{A}\hat{B}\Vert_F^2/\Vert AB\Vert_F^2$. It can be seen in Fig. \ref{fig:setwise} and \ref{fig:independent} that the empirical errors obtained by using the optimal sampling distributions have better approximations than the ones obtained by using uniform distributions. Note that in Table \ref{table1}, we can observe that in most cases, coded set-wise sampling has better approximations than coded independent sampling for the same recovery threshold. This is due to the fact that it is possible for the master to sample same submatrices multiple times when using the coded independent sampling scheme. While in coded set-wise sampling, the master always samples fresh submatrices. Furthermore, the errors of coded set-wise sampling always go to zero when $s=m$ as it is equivalent to performing the exact computation of $AB$. 

\section{Conclusion \label{sec:conclusion}}
In this paper, we studied the problem of approximate coded matrix multiplication. We presented two novel coded sampling schemes where a subset of columns/rows is sampled from the matrices. The sampled submatrices are then encoded using MatDot codes. The results reveal an interesting tradeoff between recovery threshold and approximation error.  Generalizing these ideas for other coded computation schemes is an interesting future research direction. 

\bibliographystyle{IEEEbib}
\bibliography{Ref}

\end{document}